# Luis Mateus Rocha[†] and Johan Bollen[‡]

Los Alamos National Laboratory, [†]MS B256 [‡]MS B362
Los Alamos, NM 87545, USA
E-Mail: [†]rocha@lanl.gov or [†] rocha@santafe.edu, [‡]jbollen@lanl.gov
WWW: [†]http://www.c3.lanl.gov/~rocha ,[‡] http://www.c3.lanl.gov/~jbollen

# Biologically Motivated Distributed Designs for Adaptive Knowledge Management

## 1. HUMAN-COMPUTER INTERACTION AND BIOLOGY

We discuss how distributed designs that draw from biological network metaphors can largely improve the current state of information retrieval and knowledge management of distributed information systems. In particular, two adaptive recommendation systems named *TalkMine* and *@ApWeb* are discussed in more detail. *TalkMine* operates at the semantic level of keywords. It leads different databases to learn new and adapt existing keywords to the categories recognized by its communities of users using distributed algorithms. *@ApWeb* operates at the structural level of information resources, namely citation or hyperlink structure. It relies on collective behavior to adapt such structure to the expectations of users. *TalkMine* and *@ApWeb* are currently being implemented for the research library of the Los Alamos National Laboratory under the *Active Recommendation Project*. Together they define a biologically motivated information retrieval system, recommending simultaneously at the level of user knowledge categories expressed in keywords, and at the level of individual documents and their associations to other documents. Rather than passive information retrieval, with this system, users obtain an active, evolving interaction with information resources.

### 1.1 DISTRIBUTED INFORMATION SYSTEMS AND INFORMATION RETRIEVAL[1]

*Distributed Information Systems* (DIS)[2] refer to collections of electronic networked information resources in some kind of interaction with communities of users; examples of such systems are: the Internet, the World Wide Web, corporate intranets, databases, library information retrieval systems, etc. DIS serve large and diverse communities of users by providing access to a large set of heterogeneous electronic information resources. As the complexity and size of both user communities and information resources grows, the fundamental limitations of traditional information retrieval systems have become evident.

---

[1] This sub-section draws from on-going collaboration with Cliff Joslyn at the Los Alamos National Laboratory. Many of ideas here presented are undoubtedly due to him.

[2] The main abbreviations used in this article in addition to DIS are: IR (Information Retrieval), SA (Spreading Activation), and ARP (Active Recommendation Project).

*Information Retrieval* (IR) refers to all the methods and processes for searching relevant information out of information systems (e.g. databases) that contain extremely large numbers of documents. Traditional IR systems are based solely on *keywords* that index (semantically characterize) documents and a query language to retrieve documents from centralized databases according to these keywords. This setup leads to a number of flaws:

- *Passive Environments*. There is no genuine interaction between user and system. The user *pulls* information from a passive database and therefore needs to know how to query relevant information with appropriate keywords. Furthermore, such impersonal interfaces cannot respond to queries in a user-specific fashion because they do not keep user-specific information, or user profiles. The net result is that users must know in advance how to characterize the information they need (with keywords) before pulling it from the environment.
- *Idle Structure*. Structural relationships between documents, keywords, and IR patterns are not utilized. Different kinds of structural relationships are available, but not typically used, for different DIS: e.g. citation structure in scientific library databases, the hyperlink structure in the WWW, the clustering of keyword relationships into different meanings of keywords, temporal patterns of user retrieval, etc.
- *Fixed Semantics*. Keywords are initially provided by document authors (or publishers, librarians, and indexers), and do not necessarily reflect the evolving semantic expectations of users.
- *Isolated Information Resources*. No relationships are created and no information is exchanged among documents and/or keywords in different information resources such as databases, web sites, etc. Each resource is accessed with its own set of keywords and query language.

These flaws prevent traditional IR processes in DIS to achieve any kind of interesting coupling with users. No system-user evolution or learning can be achieved because of the following fundamental limitations:

- There is no *recommendation*. Because of passive environments and idle structure, IR systems cannot pro-actively push relevant information to its users about related topics that they may be unaware of.
- There is no *conversation* between users and information resources, between information resources, and between users. Because of passive environments and isolated information resources there is no mechanism to exchange knowledge, or crossover of relevant information.
- There is no *creativity*. Because of fixed semantics, isolated information resources, idle structure, and passive environments, there is no mechanism to recombine knowledge in different information resources to infer new categories of keywords used by different communities of users.

## 1.2 DRAWING FROM BIOLOGY

The limitations of traditional IR and DIS are even more dramatic when contrasted with biological distributed systems such as immune, neural, insect, and social networks. Biological networks function largely in a distributed manner, without recourse to central controllers, while achieving tremendous ability to respond in concerted ways to different environmental necessities. In particular, they are typically endowed with the ability to elicit appropriate responses to specific demands, to transfer and process relevant information across the network, and to adapt to a changing environment by creating novel behaviors (often from recombination of existing ones). These abilities are precisely what has been lacking in IR, in which context they become ways to surmount the recommendation, conversation and creativity limitations described above.

Biological networks effectively evolve in an open-ended manner; we would like to endow DIS with a similar open-ended capacity to evolve with their users – to achieve an open-ended semiosis with them [Rocha, 2000b]. In biology, open-ended evolution originates from the existence of material building blocks that self-organization non-linearly [e.g. Kauffman, 1993] and are combined via a specification control, such as the genetic system, which nonetheless does not precisely describe or program the dynamical outcome [Pattee, 1973, 1982; Atlan and Koppel, 1990; Rocha, 1996, 1998]. In contrast, computer systems were precisely constructed with building blocks constrained in such a way as to allow minimum dynamic self-organization and maximum programmability, which results in no inherent evolvability [Conrad, 1990]. Therefore, to attain any evolvability in current digital computer systems, we need to program in some building blocks that can be used to realize the kind of dynamical richness we encounter in biological systems.

Biological systems possess an enabling chemistry (the building blocks) leading to fluid evolvability, as the possible interactions between a biological agent and its environment are open-ended. For instance, Gordon [this volume] shows how different bio-chemical profiles of ants with different roles in their colonies may be a reflection of their embodied interaction with the environment, and not necessarily a consequence of genetic differences. The gender of the Mississippi alligator too, rather than being genetically programmed, is environmentally regulated by the temperature the eggs encounter in the nest [Goodwin, 1994]. At all levels of biological systems we find this dynamic agent-environment coupling (or embodiment [Clark, 1997]) co-existing with the specification or loose programmability of the initial conditions for arrangements of dynamic building blocks, which then self-organize to produce phenotypes, behaviors, organizations, etc.[Pattee, 1982, 1995; Rocha and Hordijk 2000] The programmability can be genetic, immune, cognitive, or social[3]. Indeed, biological systems combine a small amount of programmability with rich dynamic building blocks to produce an unbounded set of self-organizing behaviors that can be picked up by natural selection [Rocha, 2000a].

Computer systems possess the description or programmability part, what they now need is an amount of dynamic agent-environment coupling, which is distributed and therefore not under complete control from a programming center. Mitchell [this volume], describing her Copycat system, suggests that in order to construct distributed, bottom-up systems capable of solving complicated cognitive tasks that are not explicitly programmed, one needs to endow computer systems with enabling *relationship packages*. In other words, there is a need for an *enabling substrate* to achieve dynamic agent-environment couplings with a smaller degree of programmability and a higher degree of self-organization.

The inherent material dynamics that permeates biology, "comes for free" [Moreno et al, 1994] for the evolving organism. In contrast, in computer systems, since we relinquished dynamics for full programmability, we need to program in every rule that may allow building blocks to be combined, self-organized, and selected – as if setting up the laws of an artificial physics and biology [Rocha and Joslyn, 1998]. Programming in the enabling substrate is, however, very different from programming the ultimate

---

[3] Clearly these types of programmability of different levels of biological systems are quite distinct. Genetic description is much more clearly understood [Rocha, 1996, 1998], but each level of biological organization establishes its own sets of constraints which, also describe or program the accepted behavior at a given level [Pattee, 1973; Salthe, 1993]

behavior that we wish to obtain. Rather, what is programmed are the lower-level building blocks and rules to relate them, which later self-organize computationally to produce (hopefully open-ended) evolving behaviors which in turn are selected by the demands of an environment or set of tasks we wish to see resolved. The enabling relationship packages are used to combine, re-combine, and transmit building blocks to produce new behavior that is not fully pre-specified. This bottom-up design mimics the existence in biology of low programmability and high evolvability.

The success of imbuing computer systems with distributed, bottom-up, designs from biology is apparent in such areas as optimization [Holland, 1995; Mitchell, 1996], modeling and simulation of social phenomena and organizations [Lindgren, 1991; Hutchins and Hazlehurst, 1991; Richards, McKay, and Richards, 1998], computer security [Forrest, Hofmeyr and Somayaji, 1997; also see Forrest's article in this volume], Artificial Life [Langton, 1989], and even biology itself [Schuster, 1995]. We are now interested in improving the limitations of IR in DIS utilizing biologically motivated designs.

The ultimate goal of IR is to produce or recommend relevant information to users. It seems obvious that the foundation of any useful recommendation should be first and foremost based on the identification of users and subject matter. In this sense, the goal of recommendation systems can be seen as similar to that of most biological systems, in particular immune systems: to recognize agents (users) and elicit appropriate responses from components of the distributed information network. Furthermore, the information network should learn and adapt to the community of agents (users) it interacts with – its environment. Naturally, unlike immune systems, the goal is not to be hostile to external agents but rather to produce information they find relevant and desirable: users are not to be treated as pathogens!

Nevertheless, as described in section 1.1, traditional IR does not identify users and classifies subjects only with unchanging keywords. To build more flexible IR, or, more generally, biologically motivated recommendation systems, we need to design the enabling relationship substrate precisely to accommodate the identification of users and their needs, as well as the evolving subjects stored in DIS. This substrate includes:

- A means to recognize *users*.
- A means to characterize *information resources*.
- A 2-way means to exchange knowledge between users and information resources: a *conversation* process. As information resources become more and more complex, we cannot expect a simple 1-way query to work well. Instead, we need a means to combine the interests of the user with the knowledge specific to each information resource.
- *Adaptation* mechanisms. We also want DIS to adapt to their community of users, as well as to exchange and re-combine knowledge leading to evolvability and creativity.

We describe below our efforts to include these biologically motivated design requirements to achieve a useful and more natural knowledge management of DIS. Before that though, we describe other recent efforts to improve IR.

## 2. ACTIVE RECOMMENDATION SYSTEMS

New approaches to IR have been proposed to address the limitations described in section 1.1. *Active recommendation systems*, also known as *Active Collaborative Filtering* [Chislenko, 1998], *Knowledge Mining*, or *Knowledge Self-Organization* [Johnson et al, 1998] are IR systems which rely on active computational environments that interact with and adapt to their users. They effectively push relevant information to users according to previous patterns of IR or individual user profiling.

Recommendation systems are typically based on user-environment interaction mediated by intelligent agents or other decentralized components and come in two varieties [Balabanoviƒ and Shoham, 1997]:

- In *content-based* recommendation, user profiles are created based on the system's keywords. Documents are recommended to users according to their profiles and some kind of semantic metric obtained from the associations between keywords and documents.
- In *collaborative* recommendation no description of the semantics or content of documents is involved, rather recommendations are issued according to a comparison of the profiles of several users that tend to access the same documents. These user profiles are not based on keywords, but on the actual documents retrieved.

Content-based systems depend on single user profiles, and thus cannot effectively recommend documents about previously unrequested content to a specific user. Conversely, pure collaborative systems, with no content analysis, match only the profiles of users that (to a great extent) have requested exactly the same documents; for instance, different book editions or movie review web sites from different news organizations are considered distinct documents. It is clear that effective recommendation systems require aspects of both approaches.

Hybrid approaches to recommendation usually rely on software agents and a central database. The agents have two distinct roles:

1. to retrieve and collect documents from information resources into a database or router
2. to select or filter those documents retrieved that match the profile of specific users.

This is the case, for instance, of *Fab* [Balabanoviƒ and Shoham, 1997] and *Amalthaea* [Moukas and Maes, 1998]. Systems such as these clearly establish active environments which are capable of recommendation, that is, they push topics that users may have not thought of, rely on user-specific interfaces that enable user identification, and keep track of historical data of the user-DIS interaction. In the terms used above, these systems expand IR beyond passive environments and completely idle structure (they keep track of user-environment interaction).

From the picture of IR depicted in section 1, there is clearly still much more room to improve. The structure and semantics of DIS is still largely idle in these collaborative systems, as they retain their original relations. Indeed these systems can improve considerably by clustering and ranking documents according to the semantics of keyword relationships [Kannan and Vempala, 1999] or the structure of document linkage [Kleinberg, 1998]. Many data-mining and graph-theoretical improvements can and should be used to discover hidden patterns in the structure of DIS, thus achieving a much more powerful recommendation capability.

However, our goal here is to improve recommendation systems by empowering them with biologically motivated conversation and creativity dimensions as described in section 1. Particularly, we want to enable the adaptation of structure and semantics of DIS to users. For this we need to develop more active environments and move beyond fixed semantics , isolated information resources, and mostly idle structure of DIS. In the following, we describe some of the work we have been developing in this direction.

# 3. THE ACTIVE RECOMMENDATION PROJECT

The *Active Recommendation Project*[4] (ARP), part of the Library Without Walls Project, at the Research Library of the Los Alamos National Laboratory is engaged in research and development of biologically motivated designs to escape the shortcomings of traditional IR and more recent recommendation systems. As discussed in section 1.2, in order to implement any biologically motivated designs we need to define an enabling relationship substrate. In this section we describe how we define such as substrate for our information resources and users.

## 3.1 INFORMATION RESOURCES: DISTRIBUTED MEMORY

The information resources available to ARP are large databases with academic articles. These databases contain bibliographic, citation, and sometimes abstract information about academic articles. Typical databases are *SciSearch*® and *Biosis*®; the first contains articles from scientific journals from several fields collected by ISI (Institute for Scientific Indexing), while the second contains more biologically oriented publications. We do not manipulate directly the records stored in these information resources, rather, we create a repository of records which point us to documents stored in these databases.

### 3.1.1 The XML Repository

We store pointers to published documents as XML[5] records. By working with XML records, we gain the ability to change the information associated with their respective documents, which we cannot do with the proprietary databases. Indeed, the XML records should be seen more as dynamic objects rather than static documents. Not only do we gain the ability to change the original keywords and citation information from the respective documents, but also the ability to add annotations, links to other records, associations with other types of media (e.g. sound clips), etc. Furthermore, XML records can even have associated procedures to compute relevant algorithms. We can think of XML records as archival objects, "buckets" of pointers, links, data, and code, which are not affiliated with any one particular information resource, as defined by Nelson et al [1998].

By transforming records from passive documents into active objects, we start our construction of the biologically motivated enabling substrate at the lowest level of information systems: the source data. This is an essential step to set up a distributed design. In centralized systems, documents can be passive since it will be up to a higher level program to decide if a certain document is relevant or not. In contrast, in distributed systems, much of the decision-making is off-loaded to lower-level components, which need to be endowed with computing capabilities. In this sense, records become active objects that store changing information, communicate with other components, and even perform actions (run code) on the information they store.

---

[4]More information, results, and testbed available at http://www.c3.lanl.gov/~rocha/lww.

[5] eXtendable Markup Language.

### 3.1.2 The Relational Repository

From the XML record repository we can derive relational information between records and keywords and among records: the *semantics* and the *structure* respectively. This semantic and structural relational repository provides the enabling relationship packages discussed above. They define which record objects are related and how, as well as the semantic tokens (keywords) they are associated with. We can also establish how keywords relate to one another.

From the XML repository we obtain $m$ records $r_j \in \mathbf{R}$, $n$ keywords $k_i \in \mathbf{K}$ and $o$ cited documents $s_m \in \mathbf{S}$. Notice that the cited document set $\mathbf{S}$, is larger than the set of records $\mathbf{R}$, and that these sets overlap only partially, because often records cite documents that are not themselves contained in the XML repository as a record. Furthermore, the two sets are not nested, that is, neither $\mathbf{R} \subseteq \mathbf{S}$ nor $\mathbf{S} \subseteq \mathbf{R}$. For structural analysis we need to create the citation document set $\mathbf{D}$ of all the $p$ documents $d_l$ involved in a citation relation. We can also derive all database semantic information from the relationships between $\mathbf{R}$ and the set of all keywords $\mathbf{K}$. Figure 1 depicts the raw information from the relational repository. We are currently using one information resource from ISI, with data from the years of 1996 to 1999. There are 2,915,258 records and 839,297 keywords. We plan to include another information resource and previous years very soon.

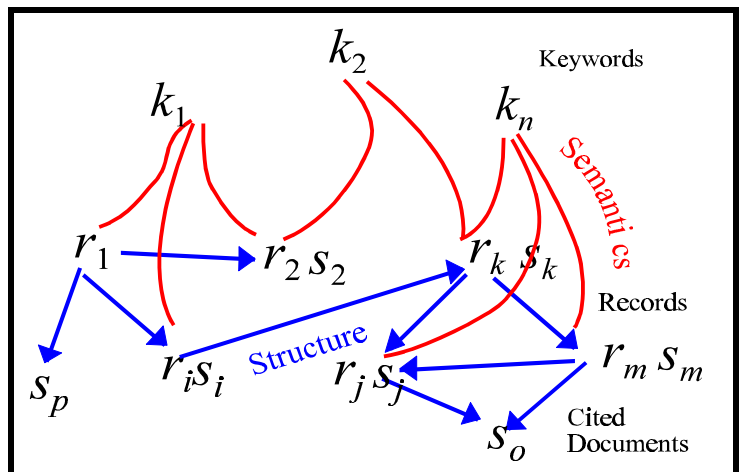

**Figure 1**: Relational Repository. The document Set D = R∪S. Some records are cited some are not. Some cited documents are records, some are not.

### 3.1.3 Structural Relations

The structure of an information resource is defined by the relations between documents in the document set $\mathbf{D}$. In academic databases these relations refer to citations, while in the World Wide Web to hyperlinks. In our case, the ISI scientific database, we work with the citation structure. Because we are working with a small interval of years, only less than half of all records (1,111,868) are an element of the set of cited documents $\mathbf{S}$, which contains 8,354,372 documents. We also discovered that many records do not participate in any citation relation (523,804), so the subset of records that participate in a citation relation is $\mathbf{R'}$ (2,391,454). The set of all documents that participate in a citation relation is $\mathbf{D} = \mathbf{R'} \cup \mathbf{S}$ (9,633,958). The citation relations are defined by the ***Citation Matrix*** $C$, a $p \times p$ matrix, of $p$ documents $d_l$ of $\mathbf{D}$. Each entry $c_{i,j}$ in the matrix is boolean and indicates whether document $d_i$ cites (1) document $d_j$ or not (0). This matrix is not symmetrical and is extremely sparse.

To discern the closeness of documents according to citation structure, we define measures of proximity between any two documents. The ***Inwards Structural Proximity Matrix*** $P^{in}$ is a square matrix of dimension $p$. For two documents $d_i$ and $d_j$, it is their direct *co-citation* [Small, 1973], that is, the number of documents that cite $d_i$ and $d_j$, over the number of documents that cite either $d_i$ or $d_j$. Documents that cite $d_i$ are referred to as ancestors of $d_i$. The Inwards Proximity varies in the unit interval and is defined by:

$$p^{in}(d_i, d_j) = \frac{\sum_{k=1}^{p}(c_{k,i} \wedge c_{k,j})}{\sum_{k=1}^{p}(c_{k,i} \vee c_{k,j})} = \frac{N_{\cap}^{in}(d_i, d_j)}{N_{\cup}^{in}(d_i, d_j)} = \frac{N_{\cap}^{in}(d_i, d_j)}{N^{in}(d_i) + N^{in}(d_j) - N_{\cap}^{in}(d_i, d_j)} \quad (1)$$

$N^{in}(d_i)$ is the number of documents that cite document $d_i$, and $N_1^{in}(d_i, d_j)$ the number of documents that cite both $d_i$ and $d_j$.

The **Outwards Structural Proximity Matrix** $P^{out}$ is a square matrix of dimension $p$. For two documents $d_i$ and $d_j$, it is their direct *bibliographic coupling* [Kessler, 1963] that is, the number of documents that both $d_i$ and $d_j$ cite, over the number of documents that either $d_i$ or $d_j$ cite. Documents that $d_i$ cites are referred to as descendants of $d_i$. The Outwards Proximity varies in the unit interval and is defined by:

$$p^{out}(d_i, d_j) = \frac{\sum_{k=1}^{p}(c_{i,k} \wedge c_{j,k})}{\sum_{k=1}^{p}(c_{i,k} \vee c_{j,k})} = \frac{N_{\cap}^{out}(d_i, d_j)}{N_{\cup}^{out}(d_i, d_j)} = \frac{N_{\cap}^{out}(d_i, d_j)}{N^{out}(d_i) + N^{out}(d_j) - N_{\cap}^{out}(d_i, d_j)} \quad (2)$$

**Table I**: 10 Most Common (stemmed) Keywords and their frequency

| Frequency | Keyword |
|---|---|
| 187705 | Cell |
| 150795 | studi |
| 149594 | system |
| 140738 | express |
| 127350 | protein |
| 124094 | model |
| 120215 | activ |
| 113740 | human |
| 112737 | rat |
| 112702 | patient |

$N^{out}(d_i)$ is the number of documents that document $d_i$ cites, and $N_1^{out}(d_i, d_j)$ the number of documents that both $d_i$ and $d_j$ cite. These very sparse directed graphs can be combined into a non-directed graph via some linear combination. From this value we can define a *neighborhood* of a document $d_i$ as the set of documents related to it with proximity greater than " 0 [0, 1]. Furthermore, we use this structural proximity information to study the relative importance of documents using singular value decomposition [Kleinberg, 1998] as well as standard clustering techniques to obtain clusters of related documents.

### 3.1.4 Semantic Relations

From the XML record repository we obtain the set of all (2,915,258) records **R** and the set of all (839,297) keywords **K**. The relations between the elements of these sets allow us to infer the semantic value of documents and the inter-relations between semantic tokens: the keywords. Naturally, semantics is ultimately only expressed in the brains of users who utilize the documents, but keywords are tokens of this ultimate expression, which we can infer from the relation between **R** and **K**. The sources of keywords are the terms authors and/or editors chose to qualify documents, as well as title words. The 10 most common keywords in our data set are listed in Table I[6].

The relations between **K** and **R** are formalized by the very sparse ***Keyword-Record Matrix*** *A*: $n \times m$ matrix, of *n* keywords $k_i$ and *m* records $r_j$. Each entry $a_{i,j}$ in the matrix is boolean and indicates whether keyword $k_i$ qualifies (1) record $r_j$ or not (0). To discern the closeness among keywords according to this relation we compute the ***Keyword Semantic Proximity Matrix*** *KSP*. It is a sparse square matrix of dimension *n*. For two keywords $k_i$ and $k_j$, it is the number of records they both qualify, over the number of records either one qualifies. Proximity varies in the unit interval, and is defined by the following equation:

$$ksp(k_i, k_j) = \frac{\sum_{k=1}^{m}(a_{i,k} \wedge a_{j,k})}{\sum_{k=1}^{m}(a_{i,k} \vee a_{j,k})} = \frac{N_{\cap}(k_i, k_j)}{N_{\cup}(k_i, k_j)} = \frac{N_{\cap}(k_i, k_j)}{N(k_i) + N(k_j) - N_{\cap}(k_i, k_j)} \qquad (3)$$

The semantic proximity calculations between two keywords, $k_i$ and $k_j$, depends on the sets of records qualified by either keyword, and the intersection of these sets. $N(k_i)$ is the number of records keyword $k_i$ qualifies, and $N_1(k_i, k_j)$ the number of records both keywords qualify. This last quantity is the number of elements in the intersection of the sets of records that each keyword qualifies. Thus, two keywords are near if they tend to qualify many of the same records. Table II presents the values of *KSP* for the 10 most common keywords, and figure 2 depicts the same information in graphical form.

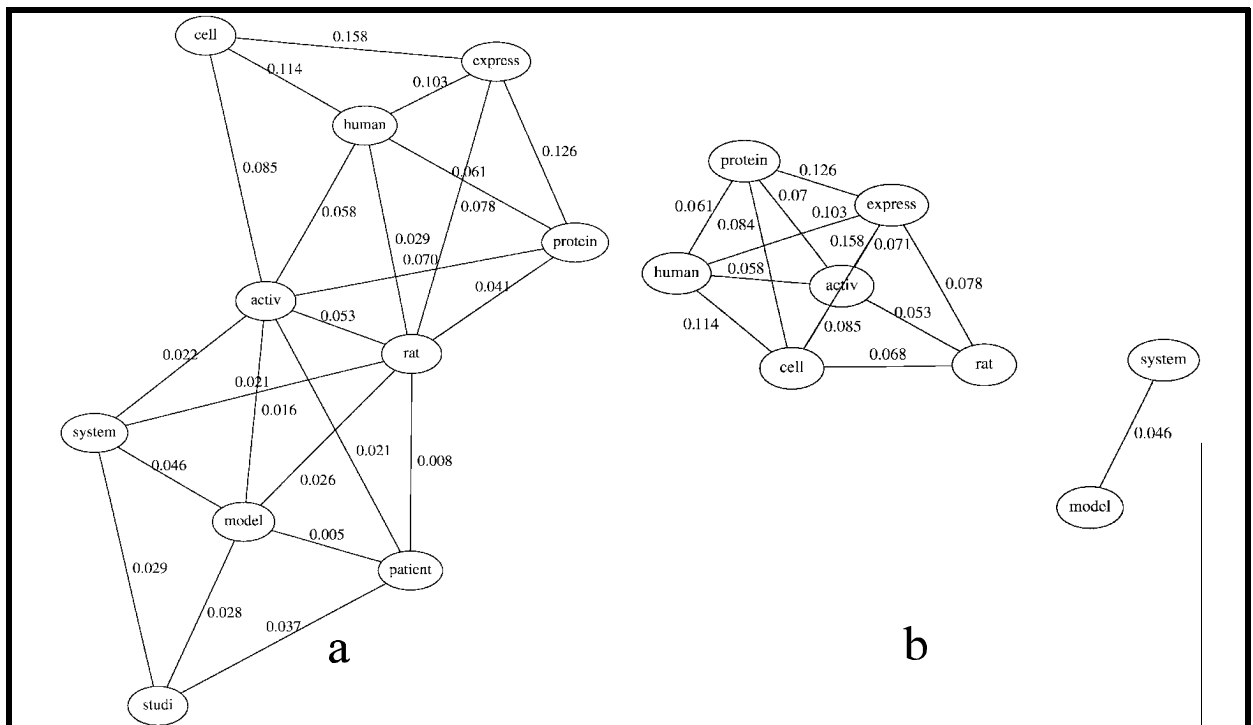

**Figure 2**: Keyword Semantic Proximity for 10 most common keywords. (a) Shows the 3 highest values for each node. (b) Shows all values higher than 0.045.

---

[6] We considered only keywords which qualify at least two records. For details about our keyword data consult: http://www.c3.lanl.gov/~rocha/lww/keywords.html.

**Table II**: Keyword Semantic Proximity for 10 most frequent keywords

|         | cell  | studi | system | express | protein | model | activ | human | rat   | patient |
|---------|-------|-------|--------|---------|---------|-------|-------|-------|-------|---------|
| cell    | 1.000 | 0.022 | 0.019  | 0.158   | 0.084   | 0.017 | 0.085 | 0.114 | 0.068 | 0.032   |
| studi   | 0.022 | 1.000 | 0.029  | 0.013   | 0.017   | 0.028 | 0.020 | 0.020 | 0.020 | 0.037   |
| system  | 0.019 | 0.029 | 1.000  | 0.020   | 0.017   | 0.046 | 0.022 | 0.014 | 0.021 | 0.014   |
| express | 0.158 | 0.013 | 0.020  | 1.000   | 0.126   | 0.011 | 0.071 | 0.103 | 0.078 | 0.020   |
| protein | 0.084 | 0.017 | 0.017  | 0.126   | 1.000   | 0.013 | 0.070 | 0.061 | 0.041 | 0.014   |
| model   | 0.017 | 0.028 | 0.046  | 0.011   | 0.013   | 1.000 | 0.016 | 0.016 | 0.026 | 0.005   |
| activ   | 0.085 | 0.020 | 0.022  | 0.071   | 0.070   | 0.016 | 1.000 | 0.058 | 0.053 | 0.021   |
| human   | 0.114 | 0.020 | 0.014  | 0.103   | 0.061   | 0.016 | 0.058 | 1.000 | 0.029 | 0.021   |
| rat     | 0.068 | 0.020 | 0.021  | 0.078   | 0.041   | 0.026 | 0.053 | 0.029 | 1.000 | 0.008   |
| patient | 0.032 | 0.037 | 0.014  | 0.020   | 0.014   | 0.005 | 0.021 | 0.021 | 0.008 | 1.000   |

Conversely, to discern the closeness of records according to relation *A*, we compute the ***Record Semantic Proximity Matrix RSP***. It is a sparse square matrix of dimension *m*. For two records $r_i$ and $r_j$, it is the number of keywords that qualify both, over the number of keywords that qualify either one. It varies in the unit interval, and it is defined by the following equation:

$$rsp(r_i, r_j) = \frac{\sum_{k=1}^{m} (a_{k,i} \wedge a_{k,j})}{\sum_{k=1}^{m} (a_{k,i} \vee a_{k,j})} = \frac{N_\cap(r_i, r_j)}{N_\cup(r_i, r_j)} = \frac{N_\cap(r_i, r_j)}{N(r_i) + N(r_j) - N_\cap(r_i, r_j)} \quad (4)$$

The semantic proximity calculations between two records, $r_i$ and $r_j$, depends on the sets of keywords qualifying either record, and the intersection of these sets. $N(r_i)$ is the number of keywords that qualify record $r_i$, and $N_\cap(r_i, r_j)$ the number of keywords that qualify both records. Thus, two records are near if they tend to be qualified by many of the same keywords.

From the inverse of these very sparse matrices we can obtain a measure of distance between keywords and between records. These distances are not Euclidean metrics because they do not observe the triangle inequality. This means that the shortest distance between two keywords or records may not be the direct link but rather an indirect pathway. Such measures of distance are referred to as semi-metrics [Galvin and Shore, 1991]. We are currently investigating if the characteristics of metricity can function as an indication of related semantic topics. The semantic side of the relational repository also allows us to conduct other IR techniques such as Latent Semantic Indexing [Berry et al, 1995; Kannan and Vempala, 1999] as well as semantic proximity clustering.

### 3.1.5 Knowledge Contexts

Each information resource (e.g. a database) is characterized by the relational information described in 3.1.2 through 3.1.4, which is obtained from the record objects of 3.1.1. The collection of this relational information associated with an information resource is an expression of the particular knowledge it conveys to its community of users. Notice that most information resources share a very large set of keywords and documents pointed to by records. However, these are organized differently in each resource, leading to different collections of relational information. Indeed, each resource is tailored to a particular community of users, with a distinct history of utilization and deployment of information by its authors and users. The same keywords will be related differently in different resources. Therefore, we refer to the relational information of each information resource as a *Knowledge Context* (figure 3).

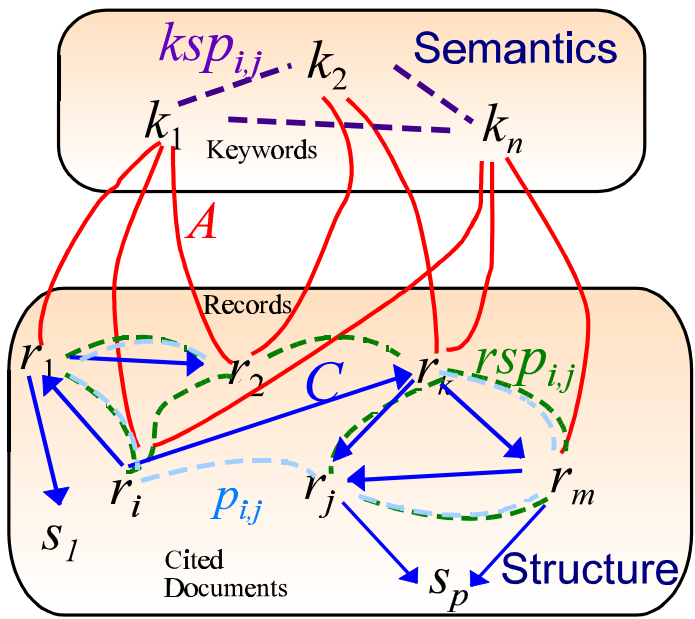

**Figure 3**: Generic Knowledge Context, with structure and semantic levels, of an information resource.

With this name we do not mean to imply that such computational structures possess cognitive abilities. Rather, we note that the way records are organized in information resources is an expression of the knowledge of its community of users. Records and keywords are only tokens of the knowledge that is ultimately expressed in the brains of users. A knowledge context simply mirrors the collective knowledge relations and distinctions of a community of users.

Notice that none of the proximity relations that define a knowledge context exist explicitly in traditional databases. Building this infrastructure is essential as an enabling relationship substrate for biologically motivated designs such as those described below in sections 4 and 5. Our object records and the proximity relations between them function as a metaphor for the material components and allowable interactions among components of biological distributed systems. Based on this substrate, we can now move to adaptive biological designs.

## 3.2 USERS

The information resources and respective knowledge contexts interact with users whose behavior we will use below to adapt the associative knowledge stored in the proximity measures. But before discussing this interaction, we need to define the capabilities of users: our agents. The following capabilities are implemented in enhanced "browsers" or centralized services that users have access to.

1. *Present interests* described by a set of keywords $\{k_1, \flat, k_i\}$.
2. *History of IR*. This history is also organized as a knowledge context as described in 3.1.5, containing the records the user has previously accessed, the keywords associated with them, as well as the structure of this set of records. This way, we treat users themselves as information resources with their own specific knowledge context defined by its own proximity information.
3. *Communication Protocol*. Users need a 2-way means to communicate with other information resources in order to retrieve relevant information, and to send signals leading to changes in all parties involved in the exchange.

The collective interaction of users defined by these capabilities, and a set of knowledge contexts from information resources of a DIS is depicted in figure 4. The knowledge contexts defined for information resources and users establish the necessary enabling substrate to set up biologically motivated designs which we describe in sections 4 and 5.

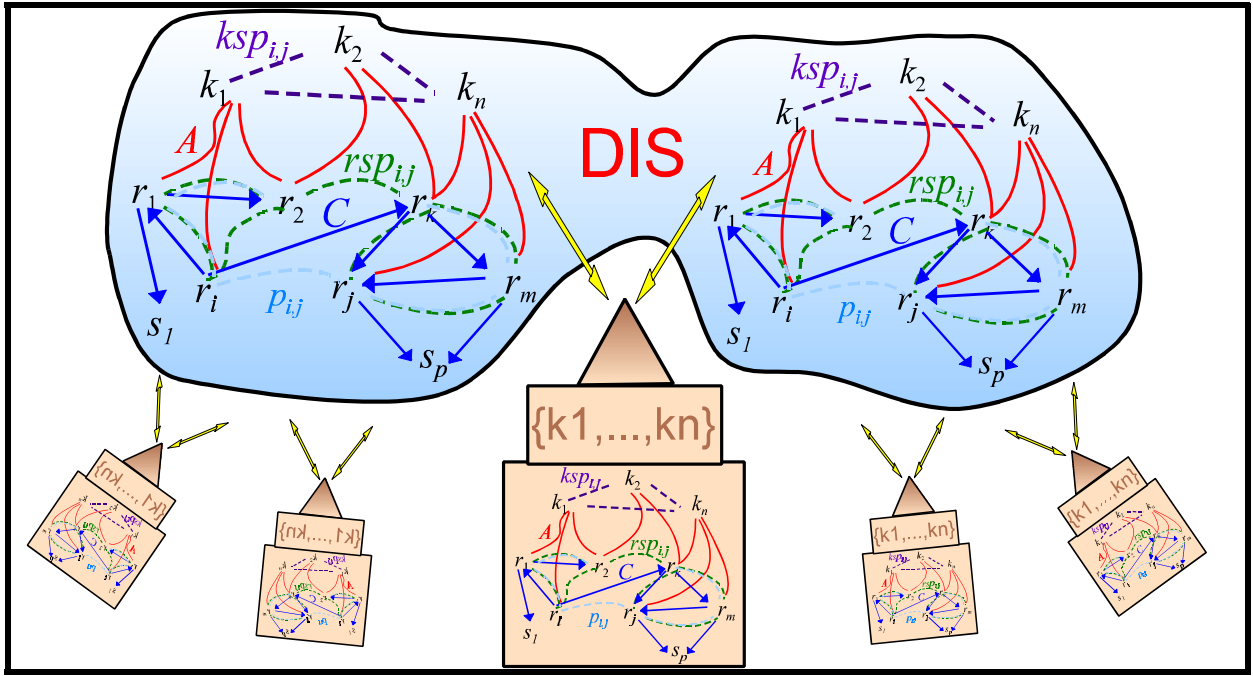

**Figure 4**: A collection of users interacts with two knowledge contexts of a DIS.

## 4. *TalkMine*: CATEGORIZATION THROUGH CONVERSATION IN DIS

Given the enabling substrate defined in section 3, to accomplish the goals expressed in section 1, we need a mechanism to enable the communication between users/agents and information resources, leading to information exchange, adaptation and recombination. *TalkMine* is a system designed especially for that. It is both a content-based and collaborative recommendation system based on a model of cognitive categories [Rocha, 1999], which are created from the conversation between users and information resources and used to re-combine knowledge as well as adapt it to users [Rocha, 2000b].

### 4.1 THE DISTRIBUTED MEMORY STRUCTURE

The proximity information of knowledge contexts is abstracted from the record-keyword ($A$) and record-record ($C$) relations and is not stored as such in the record repository. There is a parallel here to connectionist devices. Clark [1993] proposed that connectionist memory devices work by producing metrics that relate the knowledge they store (our enabling substrate). These metrics and the knowledge tokens they relate are not stored locally in the nodes of s connectionist network, but rather non-linearly superposed over its weights [van Gelder, 1991].

Our knowledge context is not a connectionist structure in a strong sense since keywords and records can be identified in particular nodes in the network. However, the same keyword qualifies many records, the same record is qualified by many keywords, and the same record typically is engaged in a citation relation with many other records. Losing or adding a few records or keywords does not affect significantly the derived semantic and structural proximity measures (as defined in section 3) of a large network. In this sense, the knowledge conveyed by such proximity measures is distributed over the entire network of records and keywords in a highly redundant manner, as required of sparse distributed memory models [Kanerva, 1988]. Below we discuss how such distributed knowledge adapts to users (the environment) with Hebbian type learning.

In the *TalkMine* model, we use the keyword semantic proximity measure (eq. 3) from the knowledge context, which we regard as the long-term memory banks of an information resource. This proximity measure is unique, reflecting the semantic relationships obtained from the set of records stored, which in turn echo the knowledge of the resource's community of users and authors. Because we use a keywords proximity, *TalkMine* is a content-based recommendation system (section 2). Next we describe how it is also collaborative by integrating the user patterns of IR.

## 4.2 SHORT-TERM CATEGORIZATION THROUGH CONVERSATION

*TalkMine* uses a set structure named *evidence set* [Rocha 1997a, 1997b, 1999], an extension of a fuzzy set [Zadeh, 1965], as a model of cognitive categories. Evidence sets are used to quantify the relative interest of users in each of the available knowledge contexts from several information resources. *TalkMine* is based on a question-answering process that integrates the user's present interests (a set of keywords) with the long-term distributed memory of the intervening knowledge contexts (including the users's). In a sense, this is done by projecting the user's interests onto the keyword proximity measures (eq. 3) of the available information resources. The result of this nonlinear integration is a category, implemented as an evidence set of keywords. This way, each user interacts with several information resources simultaneously, engaging in a multi-way conversation process.

The conversation between user and information resources is an extension of Nakamura and Iwai's [1982] question-answering IR system (for a single information resource), using uncertainty measures [Rocha, 1997a] and the evidence set operations of intersection and union [Rocha, 1999]. The algorithm of this conversation process is defined in [Rocha, 1997b, 2000b]. It constructs neighborhood functions in the semantic distances of the intervening knowledge contexts, and integrates these into an evidence set with a question-answering process that relies on (evidence set) union and intersection operations. The questions are used to reduce the uncertainty content of intermediate evidence sets, and are answered either by the user or her associated knowledge context[7]. At the end of this process an evidence set of keywords is obtained, which we regard as a knowledge category that contains the interests of the user as "seen" by the intervening information resources.

---

[7] Users possess a browser with their IR history stored as a knowledge context (section 3.2). They can set up their browsers to respond to every question themselves or allow the browser to do it automatically given the past learned experience. Users can choose an intermediate value of answering between these two extremes.

It is important to notice that the evidence set categories constructed with the question-answering algorithm, are not stored in any location in the distributed memory. They are temporarily constructed by integration of long-term knowledge from several information resources (the enabling substrate) and the present interests of the user. These constructed categories are therefore temporary containers of knowledge nonlinearly integrated from and relevant to the user and the collection of information resources. They model Clark's [1993] "on the hoof" categories. Such "on the hoof" construction of categories, triggered by interaction with users, allows several unrelated information resources to be searched simultaneously, temporarily generating categories that are not really stored in any location.

After construction of this final category, *TalkMine* returns relevant records to the user. The records returned are those that are qualified to a high degree by many of the keywords contained in the respective evidence set. Details of the actual operations used to choose relevant records are presented in [Rocha 1999].

## 4.3 ADAPTATION OF LONG-TERM MEMORY TO USERS BY SHORT-TERM CATEGORIZATION

The final component of *TalkMine* is the adaptation of the long-term distributed memory to the community of users of this system. Given the original relations of records in information resources, the derived semantic proximity measures may fail to construct associations between keywords that their users find relevant. Furthermore, the documents pointed to by records in a given information resource do not change (e.g. scientific articles), producing a fixed semantics as discussed in section 1. In contrast, the semantics of users changes with time as new keywords and associations between keywords are constantly being created and changed. Therefore, an effective recommendation system for DIS needs to adapt its knowledge contexts to the evolving semantics of its users.

The Hebbian reinforcement scheme used to implement this adaptation is very simple: the more certain keywords are combined with each other, by often being simultaneously included in the final categories, the more the distance between them is reduced. Conversely, if certain keywords are not frequently associated with one another, the distance between them is increased [details in Rocha, 1997b, 1999, 2000b]. This implements an adaption of the distributed memory of information resources to their users according to repeated inclusion of keywords in categories constructed in conversation with users. This adaptation leads the semantic proximity measures involved to increasingly match the expectations of the community of users with whom information resources interact. In other words, the distributed memory is consensually selected by the community of users.

Furthermore, when keywords in the final category are not present in one of the information resources that are combined, they are added to the information resource that does not contain them. If the association in knowledge categories of the same keywords keeps occurring, then an information resource that did not previously contain a certain keyword will have its presence progressively strengthened, even though such a keyword does not really qualify any records in this information resource.

## 4.4 EVOLVING KNOWLEDGE SYSTEMS VIA CATEGORICAL RECOMBINATION

Besides adapting independent information resources to users, *TalkMine* implements a kind of knowledge recombination that leads to evolving knowledge systems. The short-term categories bridge together a number of possibly highly unrelated contexts, which in turn creates new keyword associations in the respective information resources that would never occur within their own limited context.

Consider the following example. Two distinct information resources (databases) are going to be searched using the system described above. One database contains records of documents (books, articles, etc) of an institution devoted to the study of computational complex adaptive systems (e.g. the library of the *Santa Fe Institute*), and the other the documents of a Philosophy of Biology department. A group of users is interested in the keywords GENETICS and NATURAL SELECTION. If they were to conduct this search a number of times, due to their own interests and history, the final category obtained would certainly contain other keywords such as ADAPTIVE COMPUTATION, GENETIC ALGORITHMS, etc. Let us assume that the keyword GENETIC ALGORITHMS does not initially exist in the Philosophy of Biology digital library. After these users conduct this search a number of times, the keyword GENETIC ALGORITHMS is created in this database, even though it does not contain any records about this topic. However, with these users' continuing to perform this search over and over again, the keyword GENETIC ALGORITHMS becomes highly associated with GENETICS and NATURAL SELECTION, introducing a new perspective of these keywords. From this point on, users of the Philosophy of Biology library, by entering the keyword GENETIC ALGORITHMS would have their own data retrieval system point them to other information resources such as the library of the *Santa Fe Institute* or/and recommend documents ranging from "The Origin of Species" to treatises on Neo-Darwinism which in the meantime would have become associated with GENETIC ALGORITHMS – at which point they might rethink external access to their networked database!

*TalkMine*'s learned categories, implemented as evidence sets, integrate the knowledge of a set of information resources with user interests through conversation. This integration is in effect a temporary recombination of knowledge. If many users tend to produce similar categories, this recombination becomes fixed in the long-term distributed memory. Therefore, categorization functions as a recombination mechanism to obtain new knowledge, which can become fixed via adaptation. This is in effect a variation and selection mechanism: short-term categories provide variation of stored knowledge, while selection is implemented by the community of users. High fitness of a category corresponds to many users producing similar categories from their conversations with networked information resources. In this sense, short-term categorization not only adapts existing information resources to users, but effectively creates new knowledge in different, otherwise independent, information resources, solely by virtue of its temporary construction of categories. This open-ended semiosis of *TalkMine* is discussed in [Rocha, 2000b].

There are obvious parallels between the user-driven evolution of knowledge systems achieved by *TalkMine* and social insect models [Heylighen , 1999]. In a sense, the knowledge categories that users create function as insect trails, the more similar categories are created the more users will be attracted to them because of reinforced proximity values among their constituent keywords. In the IR world, we refer to this organization design as collaborative recommendation. As we can see, *TalkMine* is both content-based (it organizes keyword proximity measures) and collaborative (its organization is driven by user interaction). Notice however, that *TalkMine* by itself does not adapt the structure of knowledge contexts – it works solely on the semantic proximity measures. In section 5 we present another collaborative system we use in ARP to adapt the citation structure of DIS starting from *TalkMine*'s user derived keyword categories.

## 5. ADAPTIVE ASSOCIATION AND SPREADING ACTIVATION

In section 3 we arrived at a general relational structure for information resources: a knowledge context. The relations of this structure are obtained from information contained in published documents, such as keywords and citations. Notice that this information is designed by the authors and editors of these documents. Therefore, the knowledge contexts we obtain reflect the associations that authors and editors deemed relevant. The quality of these relational structures rests then on the assumption that authors and editors generate logical semantic (keywords) and structural (citation) associations with their documents. Indeed, these associations might be different from those that users may find relevant.

In section 4 we described the *TalkMine* system which in time can precisely adapt the *semantics* (the keyword proximity measures) of the original knowledge contexts to particular communities of users. In other words, *TalkMine* recommends documents according to the original knowledge derived from author and editor keyword associations, but it changes this original knowledge base according to user choices leading to an on-going evolution of the knowledge bases accessed. In this section, we describe a system used in ARP to adapt the original (e.g. citation) *structure* of knowledge contexts to users. This system, named @*ApWeb*, provides another biologically motivated means of obtaining an enabling relationship substrate on which recommendation can be issued. We also describe how another biologically motivated algorithm, *Spreading Activation* (SA), can be used for IR of the original knowledge contexts or those obtained from @*ApWeb* and *TalkMine*.

## 5.1 *@ApWeb*

*@ApWeb* is an adaptive system of obtaining a relational structure between documents of an information resource with user retrieval patterns. The knowledge contexts obtained from this system are user-determined, implicit, and collective. Its basic assumption is that the sequential retrieval of two records implies a measure of relevancy of the link or relation between them. The amount of use for a given relation between two records, is taken as an expression of its strength.

The background of this technique lies on the organization of hypertext networks such the Web. Measuring user traversal frequencies for hyperlinks [Pitkow, 1997] in a hypertext network has been proven to be quite a successful technique not only in predicting future user link preferences [Joachims, Freitag and Mitchell, 1997] but also in the interactive shaping of the structure of existing hypertext networks [Bollen and Heylighen, 1998].

In the general description of an information resource of section 3, links refer to the structural associations between records. On the web these are hyperlinks, while in academic databases they are citations. *@ApWeb* starts from the same original structure (the Citation Matrix *C* of section 3.1.3), to compute a *traversal proximity measure T*, a square matrix of $p \times p$ documents $d_l$ of the document set **D** (see section 3), that takes values in the unit interval. This measure is then used to complement the structural proximity measures (eq. 1 and 2) and the record semantic proximity (eq. 4) in a recommendation process that in this way issues recommendations according to both user patterns and author defined relations.

The original *@ApWeb* experiments [Bollen and Heylighen, 1996], were conducted on a randomly initiated structure that adapted to users in real time. In these experiments, users accessing a web page were initially shown a random set of possible links. As the adaptation process took hold, the set of links given to a user of a web page adapted to the expectations of users. Here, we describe a design[8] that relies on a fixed structure (*C*) to produce traversal proximity. In this case, when users retrieve a document, they are always shown the same set of related (by *C*) documents which they can also retrieve, namely the documents cited by the first.

---

[8] More details in [Bollen, Vandesompel, and Rocha, 1999; Bollen and Vandesompel, 2000].

Only posteriorly do we use the obtained traversal proximity to recommend relevant documents not necessarily associated in *C*.

The algorithm for obtaining *T* is the following[9]:

1. Initialize *T*. $\forall_{i,j=1}^{p} t_{i,j} = 0$.
2. Obtain the *n* user paths. A user path is the 3-tuple $B(d_i, d_j, d_k)$ of 3 documents retrieved sequentially by the same user.
3. For each path $B(d_i, d_j, d_k)$ apply the following *learning rules*:
   a. **Frequency**: $t_{i,j} = t_{i,j} + r_{freq}$ and $t_{j,k} = t_{j,k} + r_{freq}$. This rule implements a form of Hebbian learning in that the proximity between two documents that have been retrieved sequentially by the same user is reinforced with reward $r_{freq}=1/n$.
   b. **Symmetry**: $t_{j,i} = t_{j,i} + r_{symm}$ and $t_{k,j} = t_{k,j} + r_{symm}$. This rule instantiates a partially symmetric proximity. If the proximity between $d_i$ and $d_j$ increases by $r_{freq}$ with the frequency rule, the proximity between $d_j$ and $d_i$ increases by $r_{symm} < r_{freq}$ with the symmetry rule. In our experiments we have used $r_{symm} = 0.3\ r_{freq}$.
   c. **Transitivity**: $t_{i,k} = t_{i,k} + r_{trans}$. This rule instantiates a partially transitive proximity. If the proximity values between $d_i$ and $d_j$ and $d_j$ and $d_k$ increase by $r_{freq}$ with the frequency rule, the proximity between $d_i$ and $d_k$ increases by $r_{trans} < r_{freq}$ with the transitivity rule. In our experiments we have used $r_{trans} = 0.5\ r_{freq}$.

After computing *T* for a large number of user traversals, when a document is retrieved , we can recommend more documents than those it initially cited. We describe this process in 5.2, but first let us discuss how we collect user paths.

Any recommendation system will eventually produce a list of relevant documents to users. A traditional IR system (section 1) may obtain such a list from simple keyword lexical matching, while a system like *TalkMine* obtains it from the keyword categorization process described in section 4. But ultimately, we obtain a list of documents. Once a user selects one of these documents, he begins a browsing path that we can use for *@ApWeb*. Each document selected, which we store in the object records (3.1.1), is associated with a set of other documents in the particular structure of each information resource. Initially, this set may be solely the cited documents and the derived proximity measures (eq. 1, 2, and 4). The browsing path follows this structure and is stored in logs subsequently used for *@ApWeb*.

We have used *@ApWeb* to produce a user traversal proximity for the 423 web pages comprising the *Principia Cybernetica Project* web site[10] [Bollen, Vandesompel, and Rocha, 1999], as well as producing a proximity measure for the approximately 800 academic journals of the documents in the ARP database [Bollen and Vandesompel, 2000]. In the latter case, every time a user retrieves a document, its respective journal was recorded, therefore obtaining journal traversal paths which were fed to the *@ApWeb* algorithm. The journal proximity obtained provides an associative network of related journals. Below we describe how this information is used for recommendation.

---

[9] Note that other normalization schemes are possible; the 3 leaning rules are the important part.

[10] Http://pespmc1.vub.ac.be mirrored at http://pcp.lanl.gov.

## 5.2 RECOMMENDATION WITH SPREADING ACTIVATION

The technique of spreading activation (SA) is based on a model of facilitated retrieval [Meyer and Schvaneveldt, 1971] from human memory [Collins and Loftus, 1975; Anderson, 1983] and has also been implemented for the analysis of hypertext network structure [Pirolli, Pitkow, and Rao 1996]. The model assumes that the coding format of human memory is an associative network in which the most similar memory items have strongest connections [Hinton and Anderson, 1981; Klimesch, 1994]. SA works by activating a set of cue nodes in an associative network which spreads out to all other related nodes modulated by the network connection weights. The nodes that directly or indirectly accumulate most (or above a certain threshold) activation energy are considered relevant to the set of initial cue nodes. The algorithm itself works with simple linear algebra by iteratively multiplying an activation vector of all network nodes by the matrix of associative weights between nodes [Bollen, Vandesompel, and Rocha, 1999].

This IR method is ideal for networks such as those defined by the proximity measures of section 3 (eq. 1 to 4), as well as the traversal proximity of section 5.1. Several SA utilities built on these proximity measures are available on ARP's web site[11]. The main advantage of SA is that it does not depend on keyword lexical matching, but rather it exploits associative knowledge contexts built from relevant information – our enabling substrate. In other words, SA defines a process of context-dependent, knowledge-driven IR [Bollen, Vandesompel, and Rocha, 1999].

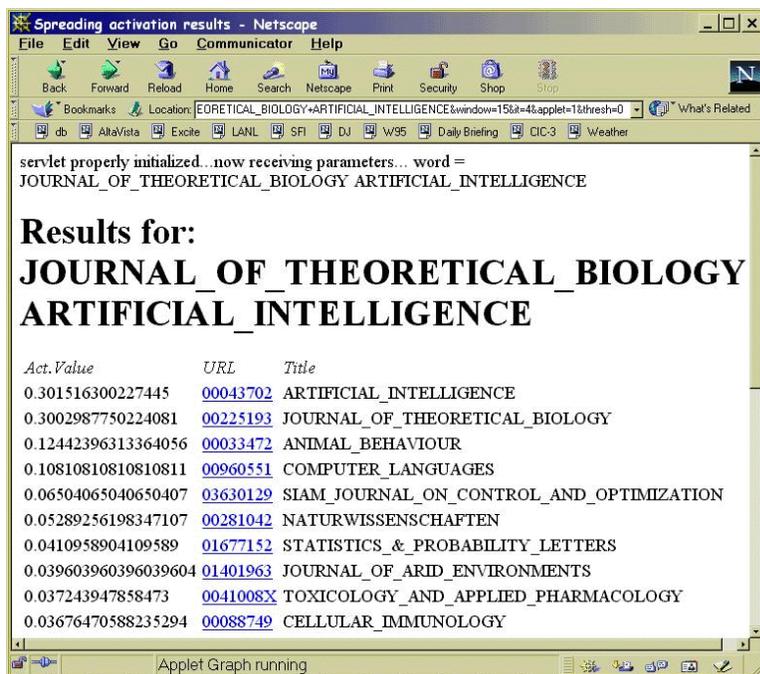

**Figure 5**: Results of spreading activation on the associative networks (traversal proximity) obtained by *@ApWeb* for journal titles.

We have applied SA to the network of journal titles obtained from *@ApWeb* described in 5.1 with very positive results [Bollen and Vandesompel, 2000]. An example of a recommendation from this utility (available in our web site) is shown in Figure 5. In this example, the initial query is JOURNAL OF THEORETICAL BIOLOGY and ARTIFICIAL INTELLIGENCE. A traditional IR search engine, such as those we typically use on the web, would perform a keyword match and return all other journal titles that include the words BIOLOGY, INTELLIGENCE, THEORETICAL, ARTIFICIAL, etc. But with the combination of SA with a *@ApWeb* organized associative network, we obtain a much more interesting recommendation of journals that do not match syntactically the query's keywords, but are in effect semantically related, e.g. STATISTICS AND PROBABILITY LETTERS, CELLULAR IMMUNOLOGY, and NATURWISSENSCHAFTEN.

---

[11] http://www.c3.lanl.gov/~rocha/lww/SA.html.

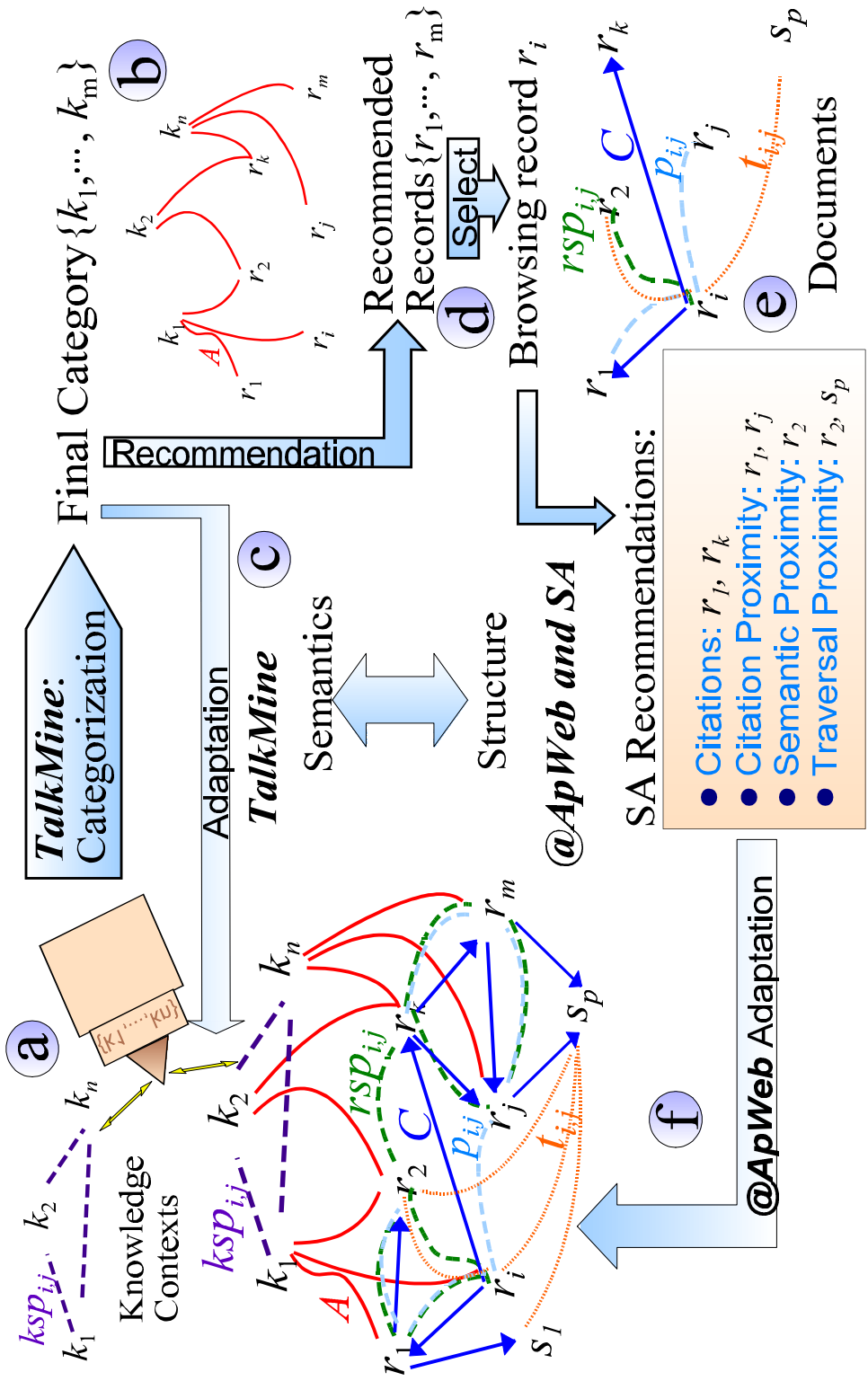

**Figure 6**: The interaction of *TalkMine* and *@ApWeb* in ARP.

Because the associative network was organized via the @*ApWeb* algorithm, these recommendations reflect the interests of the community of researchers at the Los Alamos National Laboratory. @*ApWeb* and SA define another instance of collaborative recommendation. Without access to semantic tokens such as keywords, the structure of knowledge contexts is adapted to the collective semantics of its users. Again, as discussed for *TalkMine*, this user-driven knowledge organization is based on processes similar to social insect organization. But to be able to produce such recommendation processes in DIS, we need to collect an enabling substrate of relationship packages in DIS. With the construction of the knowledge contexts of section 3, we are capable of deploying simple algorithms such as SA, while adapting them with *TalkMine* and @*ApWeb*. In the next section we describe how we integrate *TalkMine* with @*ApWeb* in ARP.

## 6. ARP AND THE FUTURE

With ARP we are interested in porting some of the evolvability of biological systems into IR distributed designs. As discussed in section 2, we also insist on recommendations systems that are both content-based and collaborative. The two systems described in sections 4 (*TalkMine*) and 5 (@*ApWeb*) operate at distinct levels of the knowledge contexts of information resources: semantics and structure respectively. This allows users to search information resources with keywords, while adapting both keywords and structural links. Their interaction in ARP is depicted in figure 6, which we describe here.

Users search several information resources using *TalkMine* to interact with their knowledge contexts (a). Users express their interests in terms of a set of keywords, which *TalkMine* integrates with the keyword proximity measures of the information resources, utilizing the question-answering categorization process described in section 4. This leads to a final category representing the interests of the user. The category is used to recommend a set records (b) and to adapt the keyword proximity measures of the intervening parties (c). The user then selects a record from the recommended set (d). The document stored in this record is shown to the user, together with a recommendation of other related documents (e). This recommendation implemented with SA (Section 5.2) utilizes the citation structure, the structural proximity measures (eq. 1, 2), the record semantic proximity measure (eq. 4), and the traversal proximity from @*ApWeb* (section 5.1). The browsing information defined by the user's document choices, the document selection paths, are then used as the @*ApWeb* adaptation signals back to the structure of the initial information resources (f).

*TalkMine* was initially developed as a prototype application for personal computers, [Rocha, 1997b, 1999], while @*ApWeb* was designed for adaptive hypertext [Bollen and Heylighen, 1996; Bollen, Vandesompel, and Rocha, 1999]. *TalkMine* is an adaptive recommendation system which is both content-based and collaborative, while @*ApWeb* with SA is strictly collaborative. The ARP testbed, where both these systems are being integrated, tackles the flaws of information retrieval in DIS as depicted in section 1 in the following manner:

- It establishes an *active environment* of user-system interaction capable of recommending information relevant to the particular users and the expectations of the overall community of users.
- It explores *structural relationships* in the document structure with proximity measures, which are now adaptive via @*ApWeb* .
- It establishes an *evolving semantics* as keyword associations adapt to the expectations of users and new keywords are introduced from the crossover of information among multiple information resources and users with *TalkMine*.

- It establishes *linked information resources* as users can search several resources simultaneously and establish all-way information exchanges.

Therefore, *TalkMine* and *@ApWeb* in ARP overcome the limitations of information retrieval outlined in section 1:

- There is *recommendation* as the system pro-actively pushes relevant documents to users about related topics that they may have been unaware of. This is achieved because of the structural and semantic proximity measures of knowledge contexts, how they are integrated with user-specific information in the categorization and adaptation processes, and finally by the document retrieval operations of *TalkMine* and SA.
- There is *conversation* between users and information resources and among information resources (and indirectly among users) as a mechanism to exchange or crossover knowledge among then is established.
- There is *creativity* as new semantic and structural associations are set up by *TalkMine* and *@ApWeb*. The categorization process brings together knowledge from the different information resources. This not only adapts existing knowledge, but combines knowledge not locally available to individual information resources. In this sense, because of the conversation process, information resources gain *new* knowledge previously unavailable.

For all of these characteristics, ARP is establishing an open-ended human-machine symbiosis, based on biologically motivated distributed designs. This design is used in the automatic, adaptive, organization of knowledge in DIS such as library databases or the Internet, facilitating the rapid dissemination of relevant information and the discovery of new knowledge.

## ACKNOWLEDGMENTS

We would like to thank Cliff Joslyn for his involvement in ARP and the development of the research presented here, Herbert Van de Sompel for his contributions for our research on Spreading Activation, and Rick Luce for making this research possible at the Los Alamos National Laboratory.

Nelson, M.L., K. Maly, N.T. Shen, and M. Zubair [1998]."Buckets: Aggregaive, intelligent agents for publishing." *National Aeronautics and Space Administration.* TM-1998-208419.

Pattee, Howard H. (Ed.) [1973]. *Hierarchy Theory: The Challenge of Complex Systems.* George Braziller.

Pattee, Howard H. [1982]."Cell psychology: an evolutionary approach to the symbol-matter problem." *Cognition and Brain Theory.* Vol. 5, no. 4, pp. 191-200.

Pattee, Howard H. [1995]."Evolving self-reference: matter, symbols, and semantic closure." *Communication and Cognition - Artificial Intelligence.* Vol. 12, nos. 1-2 (Rocha[1995a]), pp. 9-27.

Pirolli, P., J. Pitkow, and R. Rao [1996]."Silk from a sow's ear: Extracting usable structure from the web." In: *Proceedings of CHI'96 (ACM), Human Factors in Computing Systems.* . Vancouver, Canada, April 1996. ACM.

Pitkow, J. [1997]."Proceedings of the Sixth International WWW Conference." In: . . Santa Clara, California, April 1997.

Richards, D., B.D. McKay, and W.A. Richards [1998]."Collective choice and mutual knowledge structures." *Advances in Complex Systems.* Vol. 1, pp. 221-236.

Rocha, Luis M. and Cliff Joslyn [1998]."Models of Embodied, Evolving, Semiosis in Artificial Environments." In: *Proceedings of the Virtual Worlds and Simulation Conference*. C. Landauer and K.L. Bellman (Eds.). The Society for Computer Simulation, pp. 233-238.

Rocha, Luis M. and Wim Hordijk [2000]."Representations and Emergent Symbol Systems." *Cogntive Science.* Submitted.

Rocha, Luis M. [1996]."Eigenbehavior and symbols." *Systems Research.* Vol. 13, No. 3, pp. 371-384.

Rocha, Luis M. [1997a]."Relative uncertainty and evidence sets: a constructivist framework." *International Journal of General Systems.* Vol. 26, No. 1-2, pp. 35-61.

Rocha, Luis M. [1997b]. *Evidence Sets and Contextual Genetic Algorithms: Exploring Uncertainty, Context and Embodiment in Cognitive and biological Systems.* PhD. Dissertation. State University of New York at Binghamton.

Rocha, Luis M. [1998]."Selected self-organization and the Semiotics of Evolutionary Systems." In: *Evolutionary Systems: Biological and Epistemological Perspectives on Selection and Self-Organization*. S. Salthe, G. Van de Vijver, and M. Delpos (eds.). Kluwer Academic Publishers, pp. 341-358.

Rocha, Luis M. [1999]."Evidence sets: modeling subjective categories." *International Journal of General Systems.* Vol. 27, pp. 457-494.

Rocha, Luis M. [2000a]."Syntactic Autonomy: or why there is no autonomy without symbols and how self-organizing systems might evolve them." *New York Academy of Sciences.* In press.

Rocha, Luis M. [2000b]."Adaptive Recommendation and Open-Ended Semiosis ." *International Journal of Human - Computer Studies.* (In Press).

Salthe, Stanley N. [1993]. *Development and Evolution: Complexity and Changes in Biology.* MIT Press.

Schuster, P. [1995]."Artificial life and molecular evolutionary biology." In: *Advances in Artificial Life*. F. Moran, A. Moreno, J.J. Merelo, P. Chacon. Springer, pp. 3-19.

Small, H. [1973]."Co-citation in the scientific literature: a new measure of the relationship between documents." *Journal of the American Society for Information Science.* Vol. 42, pp. 676-684.

van Gelder, Tim [1991]."What is the 'D' in 'PDP': a survey of the concept of distribution." In: *Philosophy and Connectionist Theory*. W. Ramsey et al. Lawrence Erlbaum.

Zadeh, Lofti A. [1965]."Fuzzy Sets." *Information and Control.* Vol. 8, pp. 338-353.